\documentstyle[prl,aps]{revtex}

\begin{document}
\title{Symmetric Hybrid Dynamics: A canonical formulation of coupled classical-quantum dynamics}
\author{Nuno Costa Dias\footnote{ndias@mercury.ubi.pt} \\ Jo\~{a}o Nuno Prata\footnote{jprata@mercury.ubi.pt} \\ {\it Grupo de Astrof\'{\i}sica e Cosmologia} \\ {\it Departamento de F\'{\i}sica - Universidade da Beira Interior}\\ {\it 6200 Covilh\~{a}, Portugal}}
\maketitle
\rightline{GATC-00-02}
\begin{abstract}
A canonical formulation of coupled classical-quantum dynamics is presented. The theory is named symmetric hybrid dynamics. It is proved that under some general 
conditions its predictions are consistent with the full quantum ones. Moreover symmetric hybrid dynamics displays a fully consistent canonical structure. Namely, 
it is formulated over a Lie algebra of observables, time evolution is unitary and the solution of a hybrid type Schr\"odinger equation. A quantization prescription from classical mechanics to hybrid dynamics is presented. The quantization map is a Lie algebra isomorphism. Finally some possible applications of the theory are succinctly suggested.    
\end{abstract}

\subsection*{}

\paragraph*{Introduction.}

The interaction between the space time structure and the matter/energy content of the universe is described by General Relativity. The fact that all forms of matter/energy interact with the space time structure lies at the foundations of the theory. A pertinent question is then what is the form of the gravitational interaction for quantum fields \cite{hartle,wald,halliwell1}.
In the absence of a theory of quantum gravity several attempts to describe the space time - matter/energy interaction by a semiclassical formulation of gravity have been made \cite{rosenfeld,boucher}. In such a theory the space time structure is described by a classical metric while the matter content is described by standard quantum observables.
An interesting debate has been whether these semiclassical formulations will reproduce the predictions of quantum gravity in the appropriate limit \cite{page,ford}. 

This is one of several examples where what seems to be missing is a consistent general formulation of coupled classical-quantum dynamics. That is, a theory 
derived as the appropriate limit of quantum mechanics that provides a consistent description of a general interaction between classical and quantum subsystems \cite{halliwell1,halliwell2,maddox}. 

It seems reasonable to expect that such a theory should satisfy the two following prerequisites: a) under some general conditions - concerning the initial data 
and the dynamical structure of the particular system under consideration - its predictions should be consistent with the full quantum ones, and b) the theory should display a consistent dynamical structure.  

In \cite{boucher,aleksandrov,anderson} two different proposals for a theory of coupled classical-quantum dynamics were postulated and motivated in terms of the 
thus resulting properties. It was also proved \cite{nuno2} that the theory presented in \cite{boucher,aleksandrov} can be derived as the appropriate limit of 
quantum mechanics. Unfortunately, the two proposals \cite{boucher,aleksandrov,anderson} do not display a consistent canonical structure. The theories are not 
formulated over a Lie algebra of observables and thus time evolution is problematic at several levels \cite{diosi,jones}.
In fact, it was later proved that standard classical and quantum mechanics cannot be consistently coupled \cite{salcedo1,salcedo2}.
   
Recently a new formulation of hybrid dynamics was presented by Diosi, Gisin and Strunz \cite{diosi}. The starting point of this proposal is quantum mechanics in 
the Schr\"odinger picture. Hybrid dynamics is then derived using coherent state methods under some assumptions concerning the initial data and the dynamical 
behaviour. However, it is not clear in that context whether it will display a fully developed canonical structure.

In this letter we shall present an alternative proposal for coupled classical quantum dynamics. The starting point will be the full quantum formulation (in the 
Heisenberg picture) of a general dynamical system. Using a specific order dequantization \cite{nosso1} we will be able to derive symmetric hybrid dynamics. The 
procedure automatically ensures that, firstly, the theory is consistent with quantum mechanics in the appropriate limit and secondly, it displays a standard canonical structure. 

\paragraph*{Canonical Structure.}

Let us then start by establishing the conventions and assembling some general results. Let us consider a generic $(N+M)$-dimensional quantum system with fundamental observables $(\hat q_k, 
\hat p_k)$ $(k=1, \cdots, N+M)$ or sucinctly just $\hat O_k$ $(k=1, \cdots , 2(N+M))$. These operators act on the Hilbert space ${\cal H} = {\cal H}_1 \otimes 
{\cal H}_2$. The Hilbert space ${\cal H}_1$ is spanned by the eigenvectors $|z_1, \cdots , z_N>$ of the classical sector observables $\hat O_i$ $(i=1, \cdots, 
2N)$ or $(\hat q_i, \hat p_i)$ $(i=1, \cdots, N)$. The remaining subspace ${\cal H}_2$ is spanned by the eigenvectors $|w_1, \cdots , w_M>$ of the quantum sector 
observables $\hat O_{\alpha} $ $(\alpha = 2N+1, \cdots , 2(N+M))$ or $(\hat q_{\alpha}, \hat p_{\alpha} )$ $(\alpha = N+1, \cdots , N+M)$. 

The algebra of linear operators of the quantum system is generated by the set:
\begin{equation}
\hat {\cal G} \equiv \left\{ \hat O_k, \qquad k=1, \cdots, 2(N+M) \right\}.
\end{equation}
This algebra is an infinite dimensional complex vector space. In \cite{nosso1}, we saw that a possible basis for this vector space is given by the set of completely 
symmetric operators:
\begin{equation}
\hat{{\cal B}} = \left\{ \hat O_{k_1 \cdots k_n} = \left( \hat O_{k_1} \cdots \hat O_{k_n} \right)_+; \qquad 1 \le k_1, \cdots , k_n \le 2(N + M); n \in {\cal N} 
\right\}.
\end{equation}
Since any observable from the classical sector commutes with all observables of the quantum sector, the set $\hat{{\cal B}}$ can be written as the set of 
elements of the form:
\begin{equation}
\hat{{\cal B}} = \left\{ \hat O_{i_1 \cdots i_n} \hat O_{\alpha_1 \cdots \alpha_m}= \left( \hat O_{i_1} \cdots \hat O_{i_n} \right)_+ \left( \hat O_{\alpha_1} 
\cdots \hat O_{\alpha_m} \right)_+ \right\},
\end{equation}
with $ 1 \le i_1, \cdots , i_n \le 2N ; 2N+1 \le \alpha_1, \cdots \alpha_m \le 2(N+ M); n,m \in {\cal N}$. These elements are linearly independent and generate 
all elements in the algebra $\hat{{\cal A}} ({\cal H}_1 \otimes {\cal H}_2)$. Therefore $\hat{{\cal B}}$ constitutes a basis for $\hat{{\cal A}} ({\cal H}_1 
\otimes {\cal H}_2)$.

Let us now define the symmetric half-dequantization map. The classical phase-space associated with the classical sector is denoted by $T^* M_1$.

\underline{{\bf Definition 1:}} {\bf Dequantization $V_S^{H.Q.}$}
$$
V^{H.Q.}_S: \hat{{\cal A}} ({\cal H}_1 \otimes {\cal H}_2) \to {\cal S}= \left\{f : T^*M_1 \to \hat{{\cal A}} ({\cal H}_2) \right\}.
$$
1) $V^{H.Q.}_S$ is a linear map,\\
2) $V^{H.Q.}_S (\hat 1)= 1_C \hat 1_Q $,\\
3) $V^{H.Q.}_S \left( \hat O_{i_1 \cdots i_n} \hat O_{\alpha_1 \cdots \alpha_m} \right) =  O_{i_1 \cdots i_n} \hat O_{\alpha_1 \cdots \alpha_m}$,\\
where $O_{i_1 \cdots i_n}= O_{i_1} \cdots O_{i_n} \in T^* M_1$.

The dequantization $V_S^{H.Q.}$ satisfies the following properties:
\\
1) \underline{Dequantization of the product:} Let $\hat A, \hat B \in \hat{{\cal A}}$. We have:
$V_S^{H.Q.} (\hat A \cdot \hat B) = V_S^{H.Q.} (\hat A) \otimes V_S^{H.Q.} (\hat B)$, where $\otimes$ is an extension of the well-known $*$-product 
\cite{nosso1,moyal,wigner,weyl,lee,fairlie} and is given by:
\begin{equation}
\otimes : {\cal S} \times {\cal S} \to {\cal S}; \qquad \tilde A \otimes \tilde B = \tilde A \exp \left( \frac{i 
\hbar }{2} \hat{{\cal J}} \right) \tilde B ,
\end{equation}
where $\tilde A = V_S^{H.Q.} (\hat A)$\footnote{Notice that $\tilde A$ can be seen as an element of $\hat{{\cal A}} ({\cal H}_2)$ of the form $\tilde A = \sum_n
A_n^C \hat A_n^Q$, where $A_n^C \in T^* M_1$ and $\hat A_n^Q \in \hat{{\cal A}} ({\cal H}_2)$.}, $\tilde B = V_S^{H.Q.} (\hat B)$ and 
\begin{equation}
\hat{{\cal J}} \equiv \sum_{i=1}^N \left( \frac{ {\buildrel { \leftarrow}\over\partial}}{\partial q_i} \frac{ {\buildrel {\rightarrow}\over\partial}}{\partial 
p_i} 
-  \frac{ {\buildrel {\leftarrow}\over\partial}}{\partial p_i}  \frac{{\buildrel { \rightarrow}\over\partial}}{\partial q_i} \right).
\end{equation}
The product $\otimes$ is associative, distributive with respect to the sum and has a neutral element. Therefore the set $({\cal S}, \otimes, +)$ is a complex ring.
\\
2) \underline{Dequantization of the hermitean conjugate:}
\begin{equation}
V_S^{H.Q.} (\hat A^{\dagger}) = V_S^{H.Q.} \left( \sum_{i=1}^n c_i^* \left( \prod_{j=1}^m \hat X_{ij}\right)_+^{\dagger} \right) = \left[ V_S^{H.Q.} (\hat A)
\right]^{\dagger},
\end{equation}
where we considered the expansion of $\hat A$: $\hat A = \sum_{i=1}^n c_i \prod_{j=1}^m \hat X_{ij}$ and $\hat X_{ij} \in \hat {\cal G}$.
\\
3) \underline{Dequantization of the bracket:} Using the map $V_S^{H.Q.}$, we can define a new bracket structure in ${\cal S}$. Let $\hat A, \hat B \in \hat{{\cal A}}$. We have:
$$
\left[ \left[ , \right] \right]_M : {\cal S} \times {\cal S} \to {\cal S}; \qquad  \left[ \left[V_S^{H.Q.} (\hat A)  , V_S^{H.Q.} (\hat B) \right] \right]_M = V_S^{H.Q.} \left( 
\left[\hat A, \hat B \right] \right).
$$
The explicit form of the new bracket, which is an extension of the Moyal bracket \cite{moyal}, is given by:
\begin{equation}
\begin{array}{c}
\left[ \left[\tilde A , \tilde B \right] \right]_M = \tilde A \otimes \tilde B -  \tilde B \otimes \tilde A = \tilde A \exp \left( \frac{i \hbar}{2} \hat{{\cal J}} 
\right) \tilde B - \tilde B \exp \left( \frac{i \hbar}{2} \hat{{\cal J}} \right) \tilde A =\\
\\
= \left[ \tilde A, \tilde B \right] + \frac{i \hbar}{2} \left\{ \tilde A, \tilde B \right\} - \frac{i \hbar}{2} \left\{ \tilde B, \tilde A \right\} + 
{\cal O} (\hbar^2).
\end{array}
\end{equation}
If $\tilde A, \tilde B \in {\cal A} (T^* M_1)$ -the algebra of classical sector observables-, then the bracket $\left[ \left[ , \right] \right]_M$ reduces to the Moyal bracket. Likewise, for $\tilde A, \tilde B \in \hat{{\cal A}}({\cal H}_2)$, the bracket reduces to the quantum commutator. Futhermore if we disregard the terms of order $\hbar^2$ or higher the bracket reduces to the Boucher-Traschen bracket \cite{boucher}.
The properties of the new bracket follow immediately from the properties of the full quantum one: it is antisymmetric, linear and satisfies the Jacobi identity.
Therefore, $({\cal S}, \left[ \left[, \right] \right]_M)$ is a Lie algebra. Moreover, the map $V_S^{H.Q.}$ is a Lie algebra isomorphism. A quantization prescription for symmetric hybrid dynamics is given by the {\it symmetric half quantization map} $\Lambda_S^{H.Q.}$ which is defined by $\Lambda_S^{H.Q.} \circ V_S^{total} = V_S^{H.Q}$ (where $V_S^{total}: 
\hat{{\cal A}}({\cal H}_1 \otimes {\cal H}_2) \longrightarrow {\cal A} (T^* M_1 \otimes T^* M_2)$ named total dequantization map, is the extension of the map $V_S^{H.Q}$ to the case where the entire system is dequantized \cite{nosso1}). The map $\Lambda_S^{H.Q.}$ is also a Lie algebra isomorphism.  
\\
4) \underline{Dynamical Structure:} Consider the algebraic structure of symmetric hybrid dynamics. We can easily obtain the dynamical equations of the theory. They follow
directly from dequantizing the full quantum equations. The time evolution of a general observable $\tilde A$ is thus given by:
\begin{equation}
\dot{\tilde A} = \frac{1}{i\hbar} \left[ \left[ \tilde A, \tilde H \right] \right ]_M,
\end{equation}
where $\tilde H = V_S^{H.Q.} (\hat H)$ or, alternatively, $\tilde H = \Lambda_S^{H.Q.} (H)$ is the hybrid hamiltonian. The formal solution of this equation is given by:
\begin{equation}
\tilde A (t) = \sum_{n=0}^{+ \infty} \frac{1}{n !} \left( \frac{it}{ \hbar} \right)^n \left[ \left[\tilde H , \left[ \left[ \tilde H , \cdots \left[ \left[ \tilde H, \tilde A \right] \right]_M  
\cdots   \right] \right]_M  \right] \right]_M.
\end{equation}
Alternatively, the time evolution of $\tilde A$ is generated by an unitary operator $\tilde U$:
\begin{equation}
\tilde A (t) = \tilde U (t)^{\dagger} \otimes \tilde A(0) \otimes \tilde U (t),
\end{equation}
where $\tilde U(t)$ is the solution of the hybrid Schr\"odinger equation:
\begin{equation}
i \hbar \frac{\partial \tilde U}{ \partial t} = \tilde H \otimes \tilde U, \qquad \tilde U (0) =1_C \hat 1_Q,
\end{equation}
and satisfies $\tilde U^{\dagger} = \tilde U^{-1}$. Time evolution is a canonical transformation. More generally, all unitary transformations are canonical
transformations:
\begin{equation}
\tilde U^{-1} \otimes \left[ \left[ \tilde A, \tilde B \right] \right]_M \otimes \tilde U = \left[ \left[\tilde U^{-1} \otimes   \tilde A \otimes \tilde U, 
\tilde U^{-1} \otimes \tilde B \otimes \tilde U \right] \right]_M,
\end{equation}
and so the dynamical structure of the algebra $({\cal S},\otimes )$ is invariant under unitary transformations. 

\paragraph*{Predictions.}

Let the initial data for a general hybrid system be given by the initial wave function $|\phi^Q> \in {\cal H}_2$ for the quantum sector plus a set of values $(O_i^0,\delta_i), i=1..2N$ for the classical sector (where $\delta_i$ are the classical error margins associated to the initial values $O_i^0$ of the classical sector observables $O_i$).  
The aim now is to obtain physical predictions for the time evolution of the system. 
The naive procedure would be to determine a set of eigenvectors of $\tilde{O}_k(t)=\tilde{O}_k(\hat{O}_{\alpha},O_i^0,t)$ spanning the Hilbert space 
${\cal H}_2$ (let these eigenvectors be $|b_k,m_k>$, so that $\tilde{O}_k(t)|b_k,m_k>=b_k|b_k,m_k>$,
where $b_k$ is the associated eigenvalue and $m_k$ is the degeneracy index) and
then assume that the predictions of hybrid dynamics, for the outputs of a measurement of the observable $O_k$ at the time $t$, consist of the set of values $b_k$ with associated probabilities $p(b_k)=\sum_{m_k} |<b_k,m_k|\phi^Q>|^2$.

However there is no reason to believe that such predictions are physically valid. The reason is straightforward. The {\it real physical observable} is 
$\hat{O}_k(t)=\hat{O}_k(\hat{O}_{\alpha},\hat{O}_i,t)$ obtained using the full quantum formulation of the dynamical system. $\hat{O}_k(t)$ has the eigenvectors 
$|a_k,n_k>$, where $a_k$ is the associated eigenvalue and $n_k$ is the corresponding degeneracy index. The physical predictions for the output of a measurement 
of the observable $O_k$ are given by the values $a_k$ with probability $p(a_k)=\sum_{n_k} |<a_k,n_k|\phi>|^2$ where $|\phi>$ is the total wave function describing both the classical and the quantum sectors at the initial time. Clearly there is no reason why these predictions should be the same as the ones obtained by using the operator $\tilde{O}_k$.

Our best chance will be to use the operator $\tilde{O}_k$ to obtain some knowledge (but not the complete knowledge) about the outputs of a measurement of 
$\hat{O}_k$. This can be done through a procedure similar to the one presented in \cite{nuno2}. Let us then summarize the main points of that method:  

{\bf i)} The first step is the following:
quantum and hybrid dynamics provide two alternative descriptions of the initial time configuration of the dynamical system. The true, physical, description is the quantum one given by a total wave function $|\phi>$ that we assume, for simplicity, to be of the form $|\phi>=|\phi^Q>|\phi^c>$ where $|\phi^Q>\in {\cal H}_2$ and $|\phi^c>\in {\cal H}_1$. These two wave functions describe the quantum and the classical sector initial time configurations, respectively. 
On the other hand hybrid dynamics provides only an approximate description of the initial time configuration. Concerning the quantum sector, we are given exactly the same description: $|\phi^Q>$, whereas for the classical sector, we only have the set of values $(O_i^0,\delta_i)$ available. Our main task is then to understand under which conditions the two descriptions of the classical sector are consistent. Clearly, if they are not consistent, then we can not expect hybrid dynamics to yield sensible predictions.
This problem was studied in detail in \cite{nuno1} in the context of the semiclassical limit of quantum mechanics and the result turned out to be a set of 
criteria establishing a notion of classicality. These criteria were then used in \cite{nuno2} to test the physical validity of a proposed theory of hybrid 
dynamics originally suggested in \cite{boucher,aleksandrov}.

The classicality criterion consisted of a set of conditions relating the classical and the quantum description of the initial time configuration of a general dynamical system that, when satisfied, ensure that the classical predictions be consistent (in some precise sense) with the quantum ones at all times.
Here we shall also use this classicality criterion. We thus impose that $|\phi^c>$ be first order classical (i.e. it should satisfy the first order 
classicality criterion) with respect to the classical data $(O^0_i,\delta_i)$ (notice, however, that higher order classicality criteria, imposing more stringent 
conditions on the functional form of the classical sector initial data wave function, can also be used \cite{nuno2,nuno1}). This means that $|\phi^c>$ should satisfy:
\begin{equation}
<E(\hat S_{i_n},|\phi^c>,S_{i_n})|E(\hat S_{i_n},|\phi^c>,S_{i_n})> \le \delta_{S_{in}}^2
\end{equation}
where $|E(\hat S_{i_n},|\phi^c>,S_{i_n})>$ is the error ket associated to the sequence of observables 
$S_{i_n}=O_{i_1}....O_{i_n}; 1\le i_1,..,i_n \le 2N; n\in {\cal N}$,  defined by:
\begin{equation}
|E(\hat S_{i_n},|\phi^c>,S_{i_n})>=(\hat{O}_{i_1}-O^0_{i_1})....(\hat{O}_{i_n}-O^0_{i_n})|\phi^c>,
\end{equation}
and $S_{i_n}$ is any sequence of observables $S_{i_n}=O_{i_1}....O_{i_n}$ such that:
\begin{equation}
\frac{\partial^n \tilde O_k(t)}{\partial S_{i_n}} = \frac{\partial^n \tilde O_k(t)}{\partial O_{i_1}....\partial O_{i_n}} \not= 0,
\end{equation}
for any of the observables $\tilde O_k(t), k=1..2(N+M)$. Moreover $\delta_{S_{in}}=\delta_{{i_1}}....\delta_{{i_n}}$. 

In summary, a general quantum system admits a hybrid description with initial data $(|\phi^Q>,O_i^0,\delta_i)$ if the initial data wave function $|\phi^c>$ describing the classical sector in the full quantum description admits a proper classical description. This in turn means that $|\phi^c>$ should satisfy the set of conditions (13) for all sequences satisfying (15).

{\bf ii)} In the second step of our approach, a general relation between the operators $\hat{O}_k(t)=\hat{A}$ and $\tilde{O}_k(t)=\hat{B}$ is presented 
and then used to establish a relation between the eigenvectors of $\tilde{O}_k(t)$ and the eigenvectors of $\hat{O}_k(t)$.
Let us then proceed along these lines. It can be proved \cite{nuno2,nosso1}, that for $\hat B = V_S^{H.Q.}(\hat A)=\tilde A$:
\begin{equation}
\hat{A}-\hat{B} =  \sum_{i=1}^{2N} 
\frac{\partial \hat{B}}{\partial O_i}(\hat{O}_i-O_i)
+\frac{1}{2}\sum_{i,j=1}^{2N}\frac{\partial^2 \hat{B}}
{\partial O_i \partial O_j}
(\hat{O}_i-O_i)(\hat{O}_j-O_j) +...
\end{equation}
We now construct the set of eigenstates of $\hat B$, $|\psi_k,m_k>=|b_k,m_k>|\phi^c> \in {\cal H}_1 \otimes {\cal H}_2$, 
which can be used to expand $|\phi>$, and derive the explicit form of the error ket of these states, in the representation of $\hat{A}$, around the corresponding 
eigenvalue:
\begin{equation}
\begin{array}{c}
|E(\hat{A},|\psi_k,m_k>,b_k)>=( \hat{A}-\hat{B}) |\psi_k,m_k>=
\sum_{i=1}^{2N} |E(\hat{O}_i,|\phi^c>,O_i)> 
\frac{\partial \hat{B}}{\partial O_i} |b_k,m_k> +\\
\\
+ \frac{1}{2}\sum_{i,j=1}^{2N} |E(\hat{O}_i,\hat{O}_j,|\phi^c>,O_i,O_j)> 
\frac{\partial^2 \hat{B}}{\partial O_i \partial O_j}|b_k,m_k> +...
\end{array}
\end{equation}
Using this quantity it is straightfoward to prove that $\Delta(\hat{A},|\psi_k,m_k>,b_k,p)=
<E(\hat{A},|\psi_k,m_k>,b_k)|E(\hat{A},|\psi_k,m_k>,b_k)>^{1/2}/(1-p)^{1/2}$ (where  $0\le p <1$ is a probability) satisfy:
\begin{equation}
\Delta(\hat{A},|\psi_k,m_k>,b_k,p)
\le 
\frac{1}{(1-p)^{1/2}} \sum_{i=1}^{2N}   
|<b_k,m_k|\frac{\partial \hat{B}^{\dagger}}{\partial O_i}
\frac{\partial \hat{B}}{\partial O_i}|b_k,m_k>|^{1/2} \delta_{i} +...  
\end{equation}
where we explicitly used the requirement (13). $\Delta(\hat{A},|\psi_k,m_k>,b_k,p)$ is named the spread of the state $|\psi_k,m_k>$ in the representation of $\hat{A}$ and can be used to study the properties of $|\psi_k,m_k>$ in that representation. Namely it can be proved that in this representation the state $|\psi_k,m_k>$ has at least a probability $p$ confined to the interval of eigenvalues of $\hat A$, $I=[b_k-\Delta(\hat{A},|\psi_k,m_k>,b_k,p),b_k+\Delta(\hat{A},|\psi_k,m_k>,b_k,p)]$. This result is valid for all $0 \le p < 1$. The functional form of $\Delta$ suggests that we write it as 
$\Delta(\hat{A},|\psi_k,m_k>,b_k,p)=\delta_B / (1-p)^{1/2}$ where $\delta_B=<E(\hat{A},|\psi_k,m_k>,b_k)|E(\hat{A},|\psi_k,m_k>,b_k)>^{1/2}$ is named the {\it error of} $\hat B$ and is of the size of a classical error margin.  

{\bf iii)} Using the former relation between the eigenvectors of $\hat{B}$ and the eigenvectors of $\hat{A}$ we are able to derive a relation between the 
probabilities in the representation of $\hat{B}$ and those in the representation of $\hat{A}$ and thus to obtain predictions for the outputs of a measurement of $\hat{A}$
using only the knowledge about the hybrid operator $\hat B$ and its eigenvectors $|b_k,m_k>$.
Let us then consider an arbitrary interval of eigenvalues of $\hat{A}$: $I_0=[a^0-D,a^0+D]$, $D>2\Delta(p)$ and let $P(a_k \in I_0)$ be the probability of a measurement of
the observable $\hat A$ yielding a value $a_k$ inside the interval $I_0$. The full quantum mechanical predictions are given by: 
\begin{equation}
P(a_k \in I_0)=\sum_{n_k,a_k \in I_0} |<\phi|a_k,n_k>|^2.
\end{equation}
Now $P(a_k \in I_0)$ can be expanded using the eigenvectors of $\hat{B}$. We have:
\begin{equation}
P(a_k \in I_0)=\sum_{n_k, a_k \in I_0} \left|\sum_{m_k,b_k}<\phi|\psi_k,m_k><\psi_k,m_k|a_k,n_k> 
\right|^2 ,
\end{equation}
where we used the fact that $|\phi>=\sum_{b_k,m_k} <\psi_k,m_k|\phi>|\psi_k,m_k>$.
Using the properties of the states $|\psi_k,m_k>$ in the representation of $\hat{A}$
we can finally prove, through a quite long calculation, that:
\begin{equation}
1-\{P(b_k \notin I_{min})^{1/2}+(1-p)^{1/4}\}^2
\le P(a_k \in I_0) 
\le \{P(b_k \in I_{max})^{1/2}+(1-p)^{1/4}\}^2
\end{equation}
where $I_{min}=[a^0-(D-2\Delta),a^0+(D-2\Delta)]$, $I_{max}=
[a^0-(D+2\Delta),a^0+(D+2\Delta)]$ and $\Delta=\Delta(\hat{A},
|\psi_k,m_k>,b_k,p)$ is given by (18).
Notice that we can make $(1-p)$ as small as desired. However, this will affect the value of $\Delta$ and thus the range of the intervals $I_{min}$ and $I_{max}$. 

To see this explicitly let us make $p=0.99$. $\Delta$ in (18) is given by $\Delta = 10 \delta_B$ where $\delta_B=<E(\hat{A},|\psi_k,m_k>,b_k)|E(\hat{A},
|\psi_k,m_k>,b_k)>^{1/2}$ is of the order of magnitude of the classical error margins. Using this value of the spread we can state that, in the representation of 
$\hat A$, the eigenvalue of $\hat B$, $|\psi_k,m_k>$ has at least 99\% of its probability distribution confined to the interval $[b_k-\Delta,b_k+\Delta]$. The predictions 
for the outputs of a measurement of $\hat{A}$ are then:
\begin{equation}
\begin{array}{c}
1-(P(b_k \notin I_{min})^{1/2}+0.32)^2
\le P(a_k \in I_0)
\le (P(b_k \in I_{max})^{1/2}+0.32)^2 \\
\\
\Longrightarrow \quad
P(b_k \in I_{min})-0.74
\le P(a_k \in I_0) 
\le P(b_k \in I_{max})+0.74 
\end{array}
\end{equation}
where $I_{max},I_{min}=[a_0-(D\pm 20 \delta_B),a_0+(D\pm 20\delta_B)]$. An error of $74\%$ in a prediction of a probability is, of course, inadmissible. 

Let us now increase the value of $p$ and see what happens. Let $p=0.9999$. $\Delta$ is now given by $\Delta=100 \delta_B$ and the predictions for the outputs of a measurement of $\hat A$:
\begin{equation}
\begin{array}{c}
1-(P(b_k \notin I_{min})^{1/2}+0.1)^2
\le P(a_k \in I_0) 
\le (P(b_k \in I_{max})^{1/2}+0.1)^2 \\
\\
\Longrightarrow \quad
P(b_k \in I_{min})-0.21
\le P(a_k \in I_0) 
\le P(b_k \in I_{max})+0.21 
\end{array}
\end{equation}
An error of $21\%$ is more reasonable. Notice however that the difference between the range of $I_0$ and $I_{max},I_{min}$ increased considerably - from $20 \delta_B$ to $200 \delta_B$.
We can continue to increase the value of $p$ but this will affect the value of $\Delta$ and thus it is not likely that we might obtain more precise predictions. 

We should have expected symmetric hybrid dynamics to provide imprecise predictions since the theory uses classical data to describe the initial time 
configuration of one of its sectors and classical data are imprecise in nature. However, the degree of imprecision is very large which is due to the fact that 
the conditions imposed on the classical sector initial data wave function are the least restrictive possible. It is worth pointing out that more accurate 
predictions are possible either by completely reformulating the procedure by which symmetric hybrid dynamics makes predictions for the outputs of a measurement 
of the full quantum operators \cite{nosso4} or, albeit using the former procedure, by imposing more restrictive conditions on the classical sector initial data wave function \cite{nuno2}.

To see what happens in this latter case let us assume that the classical sector initial data 
wave function satisfies the 2nd order classicality criterion.  This means that $|\phi^c>$ should satisfy the conditions (13) for all 2nd order sequences of the 
form $S_{i_n}=(S_{i^{\prime}_n},S_{i^{\prime \prime}_n})$ where $S_{i^{\prime}_n}$ and $S_{i^{\prime \prime}_n}$ are any of the 1-order sequences determined in 
(15). In this case it can be proved \cite{nuno2} that $\Delta$ in (18) is (with a good aproximation) given by $\Delta=<E|E>^{1/2}/(1-p)^{1/4}=\delta_B / (1-p)^{1/4}$ and, moreover that the new spread allows for the same statement concerning the 
confinement of the eigenvectores of $\hat B$. The predictions for the outputs of a measurement of $\hat A$ become considerably more precise:
\begin{equation}
1-\{P(b_k \notin I_{min})^{1/2}+(1-p)^{3/8} /3^{1/2} \}^2
\le P(a_k \in I_0) 
\le \{P(b_k \in I_{max})^{1/2}+(1-p)^{3/8} / 3^{1/2} \}^2
\end{equation}
Let us make, once again, $p=0.9999$. This time $\Delta=\delta_B/(1-p)^{1/4}=10\delta_B$ and:
\begin{equation}
\begin{array}{c}
1-(P(b_k \notin I_{min})^{1/2}+0.02)^2
\le P(a_k \in I_0) 
\le (P(b_k \in I_{max})^{1/2}+0.02)^2\\
\\
\Longrightarrow \quad
P(b_k \in I_{min})-0.04
\le P(a_k \in I_0) 
\le P(b_k \in I_{max})+0.04 
\end{array}
\end{equation}
That is an error of at most $4\%$, with the difference between the range of $I_0$ and $I_{min},I_{max}$ decreasing to $20\delta_B$. 
If we keep increasing the degree of the classicality conditions to be satisfied by the classical sector initial data wave function we will certainly increase 
the degree of precision of the hybrid predictions. However, we will also narrow the range of validity of hybrid dynamics to those systems with a classical subsystem satisfying higher order classicality criteria. 

\paragraph*{Example.}

To illustrate the former results let us consider the simple hybrid system described by the hamiltonian:
\begin{equation}
\tilde H = \frac{1}{2}(x^2 + q^2)+kx\hat Q
\end{equation}
where $(q,x)$ are the canonical variables of a classical harmonic oscillator with $m=w=1$, $(\hat Q,\hat P)$ are the fundamental observables of the quantum system to which the harmonic oscillator is coupled and $k$ is the coupling constant. By solving the system of equations (8) we obtain the time evolution of the fundamental observables:
\begin{eqnarray}
\tilde q (t) & = & q(0) \cos t + \{x(0)+k\hat Q(0)\} \sin t \nonumber \\
\tilde x(t) & = & -q(0) \sin t +x(0) \cos t + k \{\cos t -1 \} \hat Q(0) \nonumber \\
\tilde Q(t) & = & \hat Q(0) \nonumber \\
\tilde P(t) & = & \hat P(0) - k \{q(0) \cos t + x(0) \sin t \} - k^2 \{ \sin t -t \} \hat Q(0)
\end{eqnarray}
where $(q(0),x(0))$ and $(\hat Q(0),\hat P(0))$ are the initial time classical and quantum sector fundamental observables, respectively.
Finally the $L$-order spreads (18) ($L \in {\cal N}$) are given by:
\begin{eqnarray}
\Delta_L(\hat q, p) &=& \delta_{\tilde q} / (1-p)^{1/2L} =
\{|\cos t | \delta_q(0) + |\sin t|\delta_x(0) \}/(1-p)^{1/2L}\nonumber \\
\Delta_L(\hat x, p) &=& \delta_{\tilde x} / (1-p)^{1/2L} =
\{|\sin t | \delta_q(0) + |\cos t|\delta_x(0) \}/(1-p)^{1/2L}\nonumber \\
\Delta_L(\hat Q, p) &=&  \delta_{\tilde Q} / (1-p)^{1/2L} = 0 \nonumber \\
\Delta_L(\hat P, p) &=& \delta_{\tilde P} / (1-p)^{1/2L} =
\{|k\cos t | \delta_q(0) + |k\sin t|\delta_x(0) \}/(1-p)^{1/2L}
\end{eqnarray}
where $0\le p <1$ is a probability, $\delta_q(0),\delta_x(0)$ are the initial data classical  error margins and the results are valid up to the order of classicality $L$ of the classical sector initial data wave function $|\phi^c>$. In particular they are valid for the first and second order spreads $L=1,2$ (if the initial data wave function $|\phi^c>$ is first or second order classical) that were previously used to obtain the general predictions (21,24).

\paragraph*{Conclusions.}

A canonical formulation of coupled classical and quantum dynamics was presented. The theory satisfies an interesting set of properties. On the one hand, and most 
importantly, for a general dynamical system satisfying some general conditions concerning its initial data and dynamical behaviour, the predictions of symmetric 
hybrid dynamics are consistent with the predictions of full quantum mechanics (and, in fact, can be used to obtain full quantum predictions) and are therefore physically valid.

On the other hand, symmetric hybrid dynamics displays a fully consistent canonical structure: i) it is formulated over a Lie algebra of observables, ii) time evolution is unitary and probabilities are positive defined, iii) the theory admits a set of general canonical transformations which are generated by the action of unitary operators and iv) time evolution, in particular, is a canonical transformation.

Moreover, the limit of symmetric hybrid dynamics when the classical sector does not exist is quantum mechanics. However, one should notice that the limit when the quantum sector does not exist is not standard classical mechanics. Instead this limit is symmetric classical mechanics, an alternative theory of classical mechanics that was proposed and studied in \cite{nosso1}, and this is the reason why the results of this letter do not contradict those of \cite{salcedo1,salcedo2}. Therefore symmetric hybrid dynamics properly generalizes both quantum and symmetric classical mechanics.
Futhermore, the quantization prescription from symmetric classical mechanics to symmetric hybrid dynamics is just a Lie algebra isomorphism and is thus not plagued with ordering ambiguities.

Finally, notice that the limit of symmetric classical mechanics when $\hbar \to 0$ is just standard classical mechanics. Likewise, in this limit, the bracket 
structure and thus the dynamical structure of symmetric hybrid dynamics are just the ones originally proposed by Boucher and Traschen in \cite{boucher}. This theory was later 
proved to be consistent with quantum mechanics \cite{nuno2} under the same assumptions that were made in this letter in what concerns 
the initial data but using an approximation procedure that discarded the contributions of the terms proportional to $\hbar^2$ or smaller. The results of the 
present letter corroborate, by straightforward order of magnitude considerations, the procedure used in \cite{nuno2} and thus the validity of the Boucher-Traschen bracket as 
a possible dynamical structure (unfortunately ill behaved) for coupled classical-quantum dynamics.    

\paragraph*{Acknowledgments}

This work was partially supported by the grants 
ESO/PRO/1258/98 and CERN/P/Fis/15190/1999.

\end{document}